\documentclass{article}
\usepackage{sprocl}
\usepackage{overcite}
\usepackage{epsfig}
\usepackage{picinpar}
\usepackage{amssymb,pifont,array,rotate,rotating,xspace}
\newcommand{\etal}{et al.}

\newcommand{\myarrow}[1]{\raisebox{-0.5mm}{\Large{\ding{224}}}}
\newcommand{\BR}[1]{{\mathrm{Br}}(#1)}

\newcommand{\DO}{D$\emptyset$}
\newcommand{\ETMISS}{E_T\!\!\!\!\!\!\!/\,\,~}

\newcommand{\MW}{\ensuremath{\mathit{m}_\mathrm{\,W}}}

\newcommand{\MZ}{\ensuremath{\mathit{m}_\mathrm{Z}}}

\newcommand{\MT}{\ensuremath{\mathit{m}_\mathrm{t}}}
\newcommand{\MH}{\ensuremath{\mathit{m}_\mathrm{H}}}

\newcommand{\GW}{\ensuremath{\Gamma_\mathrm{W}}}

\newcommand{\im}[1]{\ensuremath{#1}\xspace}
\newcommand{\Pchi}{\im{{\raise5pt\hbox{$\chi$}}}}

\newcommand{\W}{\ensuremath{\mathrm{W}}}
\newcommand{\Z}{\ensuremath{\mathrm{Z}}}

\def\GeV{\ifmmode {\mathrm{\ Ge\kern -0.1em V}}\else
                   \textrm{Ge\kern -0.1em V}\fi}%
\def\MeV{\ifmmode {\mathrm{\ Me\kern -0.1em V}}\else
                   \textrm{Me\kern -0.1em V}\fi}%
\def\keV{\ifmmode {\mathrm{\ ke\kern -0.1em V}}\else
                   \textrm{ke\kern -0.1em V}\fi}%
\def\eV{\ifmmode  {\mathrm{\ e\kern -0.1em V}}\else
                   \textrm{e\kern -0.1em V}\fi}%

\newcommand{\OK}{}

\begin{document}

\title{PROPERTIES OF THE W BOSON}

\author{L. TAYLOR}

\address{Department of Physics, Northeastern University,\\
Boston, MA 02115, USA}

\vspace*{-40mm}

\begin{center}
To appear in the {\em{Proceedings of the XVIIth International Conference on}} \\
{\em{Physics in Collision}}, 1997, Ed. H. Heath, World Scientific, Singapore.
\end{center}
\vspace*{20mm}

\hspace*{\fill}{\large\bf{L3 Note 2197}}

\vspace*{20mm}

\maketitle\abstracts{
The properties of the W boson are reviewed.
Particular emphasis is placed on
recent measurements from the LEP2 and Tevatron experiments.}
  
\section{Introduction\label{sec:intro}}

The W boson was discovered\cite{WDISCUAONE,WDISCUATWO} by the UA1 
and UA2 experiments at the CERN S${\mathrm{p\bar{p}}}$S collider 
in 1983. 
Since then, its properties have been measured by UA1 and UA2, 
by the CDF\cite{CDF} and \DO\cite{DZERO} experiments 
at the FNAL Tevatron,
and more recently by 
ALEPH\cite{ALEPH}, 
DELPHI\cite{DELPHI}, 
L3\cite{LTHREE}, and
OPAL\cite{OPAL} at 
the LEP ${\mathrm{e^+e^-}}$ collider at CERN.
In this paper we describe the production and decays of W's (Sec.~\ref{sec:proddecay});
the determination of the W width (Sec.~\ref{sec:wwidth});
constraints on anomalous couplings of the W (Sec.~\ref{sec:wcouplings}); 
and measurements of the W mass (Sec.~\ref{sec:wmass}).
\section{W Production and Decay\label{sec:proddecay}}

\subsection{W Production and Decay at LEP}

\begin{figure}[!htbb]
\begin{center}
\fbox{\epsfig{file=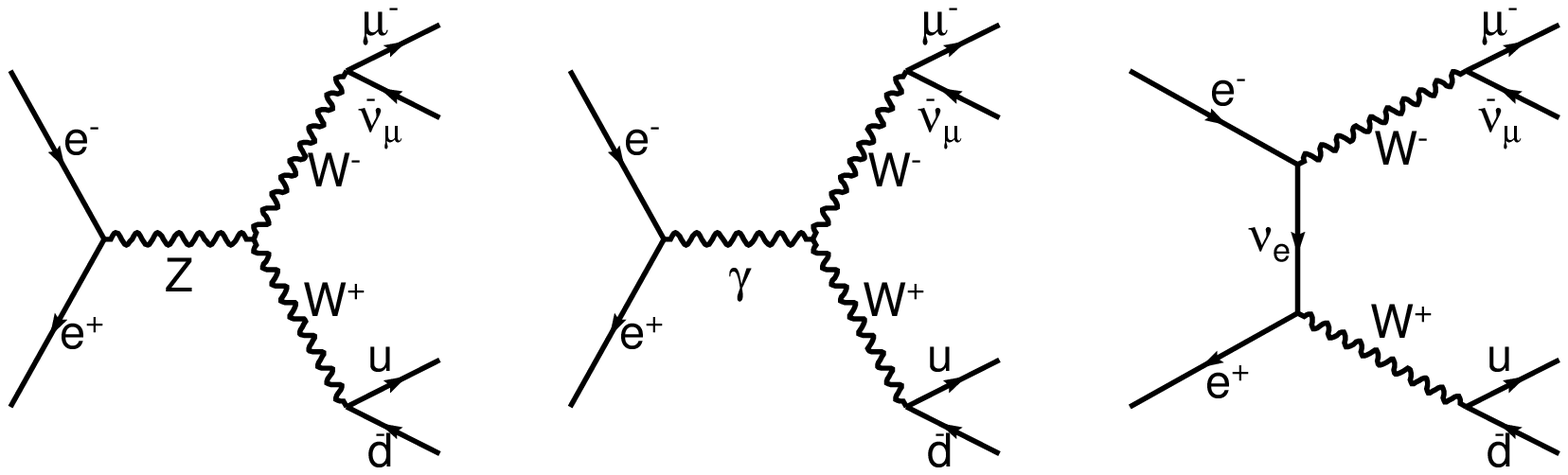,width=0.7\textwidth,clip=}} \\[1.5mm] 
\fbox{\epsfig{file=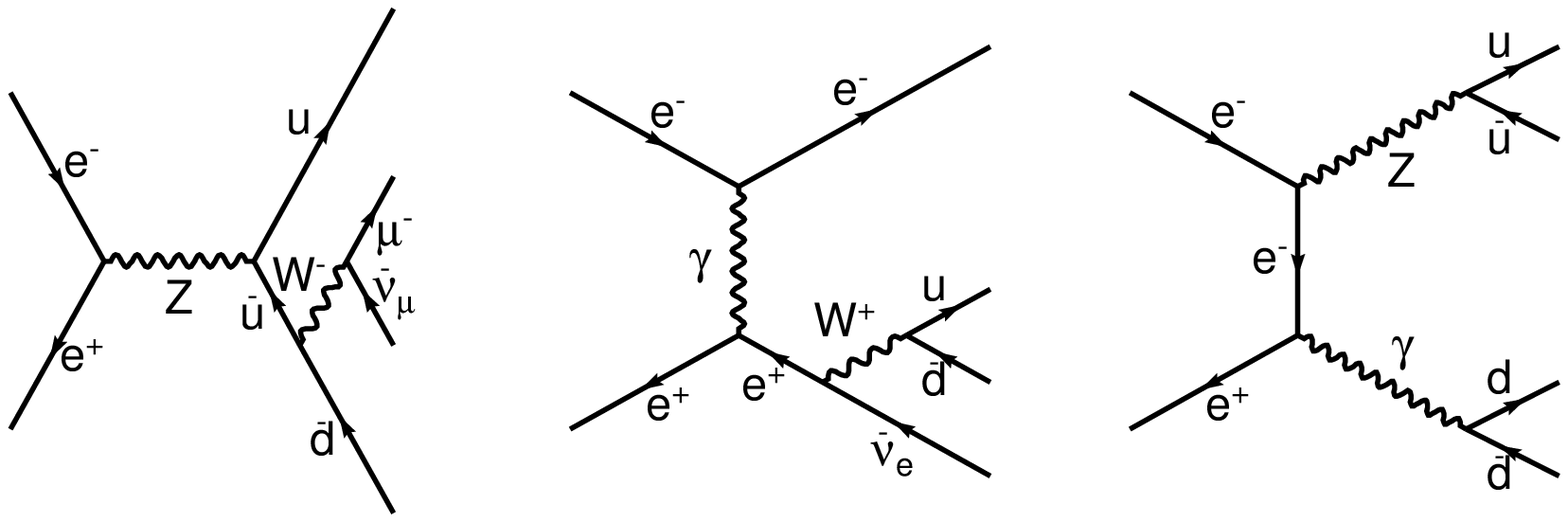,width=0.7\textwidth,clip=}}  
\end{center}
\caption{Upper: The CC03 diagrams for W-pair production in ${\mathrm{e^+e^-}}$ collisions 
         with, in this example, subsequent decays to ${\mathrm{q\bar{q}}}$
         and $\mu\nu$.
         Lower: Three important four-fermion diagrams which interfere with the 
         CC03 processes.\label{fig:wwfeynman}}
\end{figure}

W-pairs are produced at LEP II through ${\mathrm{Z}}/\gamma$ s-channel and 
neutrino t-channel processes\cite{LEPIIYELLOW} (the so-called CC03 processes), as shown in 
Fig.~\ref{fig:wwfeynman}(upper).
%
Other diagrams, such as those shown in Fig.~\ref{fig:wwfeynman}(lower) lead 
to the same final states and interfere with the CC03 processes.
The identification of W-pairs at LEP II is more difficult than that of 
Z's at LEP I due to the considerable amount of Standard Model background.
At ${\sqrt{s}}=161$\,GeV ($\sim 10$\,pb$^{-1}$/experiment) 
the backgrounds are approximately two orders of 
magnitude more than the signals.
At ${\sqrt{s}}=172$\,GeV ($\sim 10$\,pb$^{-1}$/experiment) 
the situation is about five times better due to 
the increasing W-pair cross-section. 

The main background to the channel ${\mathrm{WW\rightarrow q\bar{q}q\bar{q}}}(\gamma)$ 
comes from the ${\mathrm{e^+e^-\rightarrow q\bar{q}(\gamma)}}$ process which has 
a cross-section of approximately 150\,pb.
Signal events are selected by requiring the events to have high 
multiplicities and low missing energy.
Radiative Z return events are rejected.
The ${\mathrm{WW\rightarrow q\bar{q}\ell\nu}} (\gamma)$ samples are selected 
by requiring the event to have two high energy jets and an isolated charged lepton.
The neutrino momentum is inferred from the missing energy.
The main backgrounds come from mis-identified ${\mathrm{e^+e^-\rightarrow q\bar{q}(\gamma)}}$ 
events and from the four-fermion process  ${\mathrm{e^+e^-\rightarrow q\bar{q}\ell^+\ell^-}}$
in which one of the leptons is undetected.
The ${\mathrm{WW\rightarrow\ell\nu\ell\nu}} (\gamma)$ samples are obtained by 
requiring the presence of two charged leptons and by excluding
high multiplicity (hadronic) events.
The dominant backgrounds are from the radiative-return ${\mathrm{Z}}\rightarrow\ell^+\ell^-$,
Bhabha, and two-photon processes.

Tab.~\ref{tab:gurtu}\OK\ shows typical values for W-pair 
selection efficiencies at LEP and the resulting event sample sizes 
at ${\sqrt{s}}=161$\,GeV and ${\sqrt{s}}=172$\,GeV.
\begin{table}[!b]
{\small
\renewcommand{\arraystretch}{1.1}
\setlength{\tabcolsep}{3.7pt}
\caption{Typical values for W-pair selection efficiencies at LEP and 
         the resulting event sample sizes, 
         at ${\sqrt{s}}=161$\,GeV and ${\sqrt{s}}=172$\,GeV.\label{tab:gurtu}}
\vspace{0.4cm}
\begin{center}
\begin{tabular}{|l|c|c|c|c|} \hline
\multicolumn{1}{|c|}{WW event}                     & 
\multicolumn{2}{c|}{Selection efficiency (\%)}     & 
\multicolumn{2}{c|}{Events/experiment}  \\ \cline{2-5} 
~~sample                                & ${\sqrt{s}}=161$\,GeV    & ${\sqrt{s}}=172$\,GeV & ${\sqrt{s}}=161$\,GeV & ${\sqrt{s}}=172$\,GeV  \\ \hline
~~${\mathrm{q\bar{q}q\bar{q}}}(\gamma)$ &    $\sim$60              & 75 -- 85              &  9 -- 15              &  55 -- 65              \\
~~${\mathrm{q\bar{q}\ell\nu}} (\gamma)$ &    60 -- 80              & 60 -- 90              & 11 -- 16              &  40 -- 50              \\
~~${\mathrm{\ell\nu\ell\nu}}  (\gamma)$ &    40 -- 70              & 45 -- 80              &  2 --  6              &   5 -- 10              \\ \hline
~~Total                                 &                          &                       & 22 -- 36              &  95 -- 120             \\ \hline
\end{tabular}
\end{center}
}
\end{table}
After correcting for the measured luminosity, the detector acceptances,  
reconstruction efficiencies, and backgrounds the experiments have fitted 
their data to extract the LEP average results\cite{ALEPHWA,DELPHIWB,LTHREEWA,OPALWB,LEPPRIVW,GURTU}:
$\BR{\W\rightarrow {\mathrm{e}}\nu} = (12.0 \pm 1.9)$\%,
$\BR{\W\rightarrow  \mu        \nu} = (10.3 \pm 1.7)$\%,
$\BR{\W\rightarrow  \tau       \nu} = (10.7 \pm 2.2)$\%,
where the errors are dominated by the statistical uncertainties. 
Assuming lepton universality yields\cite{ALEPHWA,DELPHIWB,LTHREEWA,OPALWB,LEPPRIVW,GURTU}
$\BR{\W\rightarrow {\ell   \nu}}               = (11.0 \pm 0.7)$\% $(\ell={\mathrm{e}}/\mu/\tau)$
and
$\BR{\W\rightarrow {\mathrm{q\bar{q}^\prime}}} = (67.0 \pm 2.1)$\%,
in agreement with the Standard Model predictions of 
$\BR{\W\rightarrow  \ell   \nu}_{\mathrm{SM}} = 10.83$\% 
and
$\BR{\W\rightarrow {\mathrm{q\bar{q}^\prime}}}_{\mathrm{SM}} = 67.5$\%.
Assuming $\BR{\W\rightarrow \ell\nu}_{\mathrm{SM}}$, the 
cross-sections (for CC03 processes) are determined to be
$\sigma_{\mathrm{WW}}(161{\mathrm{GeV}}) = (3.67  \pm 0.42)$\,pb and
$\sigma_{\mathrm{WW}}(172{\mathrm{GeV}}) = (11.95 \pm 0.70)$\,pb.
The W mass-dependence of the cross-section close to threshold
is used to determine the W mass, as described in section~\ref{sub:wmasslep}.

\subsection{W Production and Decay at the Tevatron}

The production of W's in ${\mathrm{p\bar{p}}}$ collisions at the 
Tevatron, which occurs predominantly through quark-antiquark 
annihilation, is described elsewhere in these proceedings\cite{GUGLIELMO}.
%
%
%
%
%
%
%
%
%
%
For both CDF and \DO, candidate W events are triggered by a 
single lepton trigger and are then required to contain a charged lepton 
with transverse momentum of typically $p_T^\ell > 25$\,GeV and 
missing energy of typically $\ETMISS > 25$\,GeV.
Efficiencies for selecting leptonic W decays, including the effects of the
trigger, geometric acceptance, and selection and kinematic cuts, are
25--30\% for electrons and 5--10\% for muons\cite{DEMARTEAU}.
Typical yields for CDF(\DO) are therefore 
700(800) ${\mathrm{W\rightarrow{e}\nu}}$ decays/${\mathrm{pb}}^{-1}$ and  
150(350) ${\mathrm{W\rightarrow\mu\nu}}$ decays/${\mathrm{pb}}^{-1}$.
The corresponding backgrounds are typically 10--20\%.
These numbers correspond to the pre-selected samples and can 
change appreciably in the various analyses as more stringent
selection criteria are applied.

\begin{figure}[htb]
\begin{center}
\epsfig{file=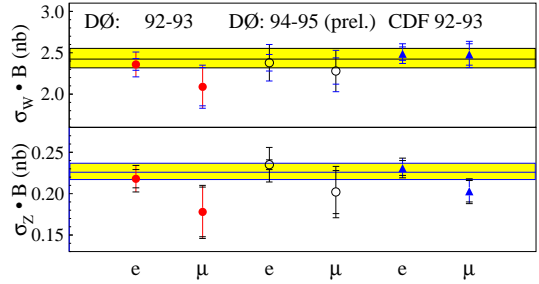,width=0.6\textwidth,clip=}
\end{center}
\caption{Summary of W and Z cross-section times leptonic branching 
         ratio measurements.
         The shaded band denotes the theoretical prediction described 
         in the text.\label{fig:wcrosssec}}
\end{figure}

CDF and \DO\ have measured the products of the cross-section and 
leptonic branching ratio, ${\sigma_\W} \cdot \BR{\W\rightarrow \ell \nu}$ and 
${\sigma_\Z} \cdot \BR{\Z\rightarrow \ell \ell}$, using the 
background-subtracted number of observed events in each channel,
corrected for acceptance, efficiency, and luminosity. 
Such measurements are a good test of our understanding of QCD 
and the Parton Distribution Functions (PDF's),
for which NLO corrections are 20\% and NNLO corrections are about 
3\%\cite{HAMBERG91,NEERVEN92}.
These calculations predict $\sigma_\W = 22.35$\,nb and $\sigma_\Z = 6.708$\,nb,
for the CTEQ2M PDF.
Using the theoretical expectation of 
$\BR{\W\rightarrow \ell \nu} = (10.84 \pm 0.02)$\%\cite{ROSNER94} 
and the LEP measurement of 
$\BR{\Z\rightarrow \ell^+ \ell^-} = (3.366 \pm 0.006)$\%\cite{PDG96},
yields expectations of 
${\sigma_\W} \cdot \BR{\W\rightarrow \ell \nu}  = (2.42^{+0.13}_{-0.11})$\,nb and
${\sigma_\Z} \cdot \BR{\Z\rightarrow \ell \ell} = (0.226^{+0.011}_{-0.009})$\,nb,
where the errors are dominated by the uncertainties in the PDF.
Fig.~\ref{fig:wcrosssec} summarises these measurements and demonstrates the
good agreement with the theoretical predictions, shown by the shaded band.
In future, experiments may ultimately use measurements of 
${\sigma_\W} \cdot \BR{\W\rightarrow \ell \nu}$ to measure
their luminosity.
 
\begin{figure}[htb]
\begin{center}
\fbox{\epsfig{file=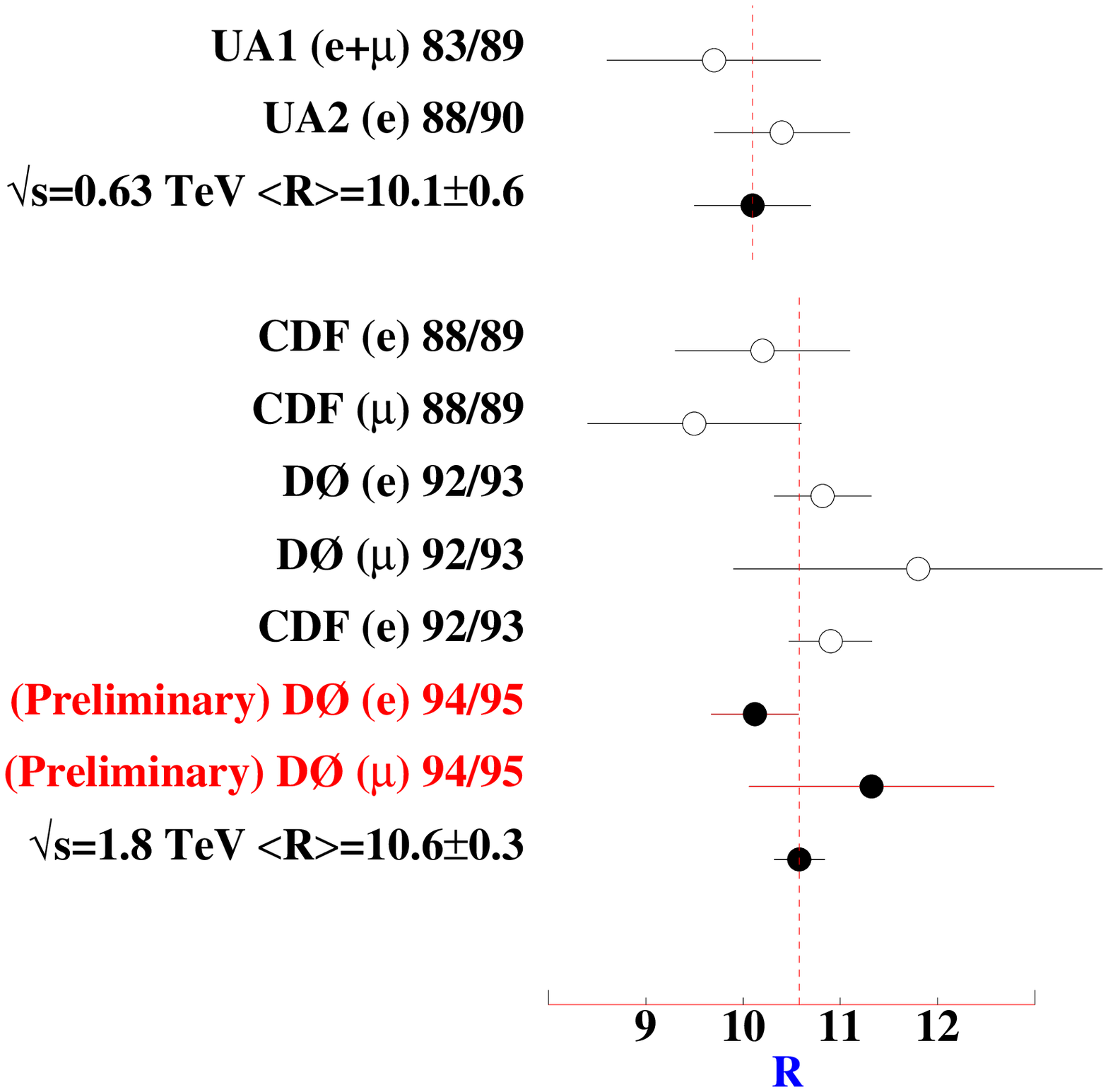,height=0.34\textheight,clip=}}~~~~~
\fbox{\epsfig{file=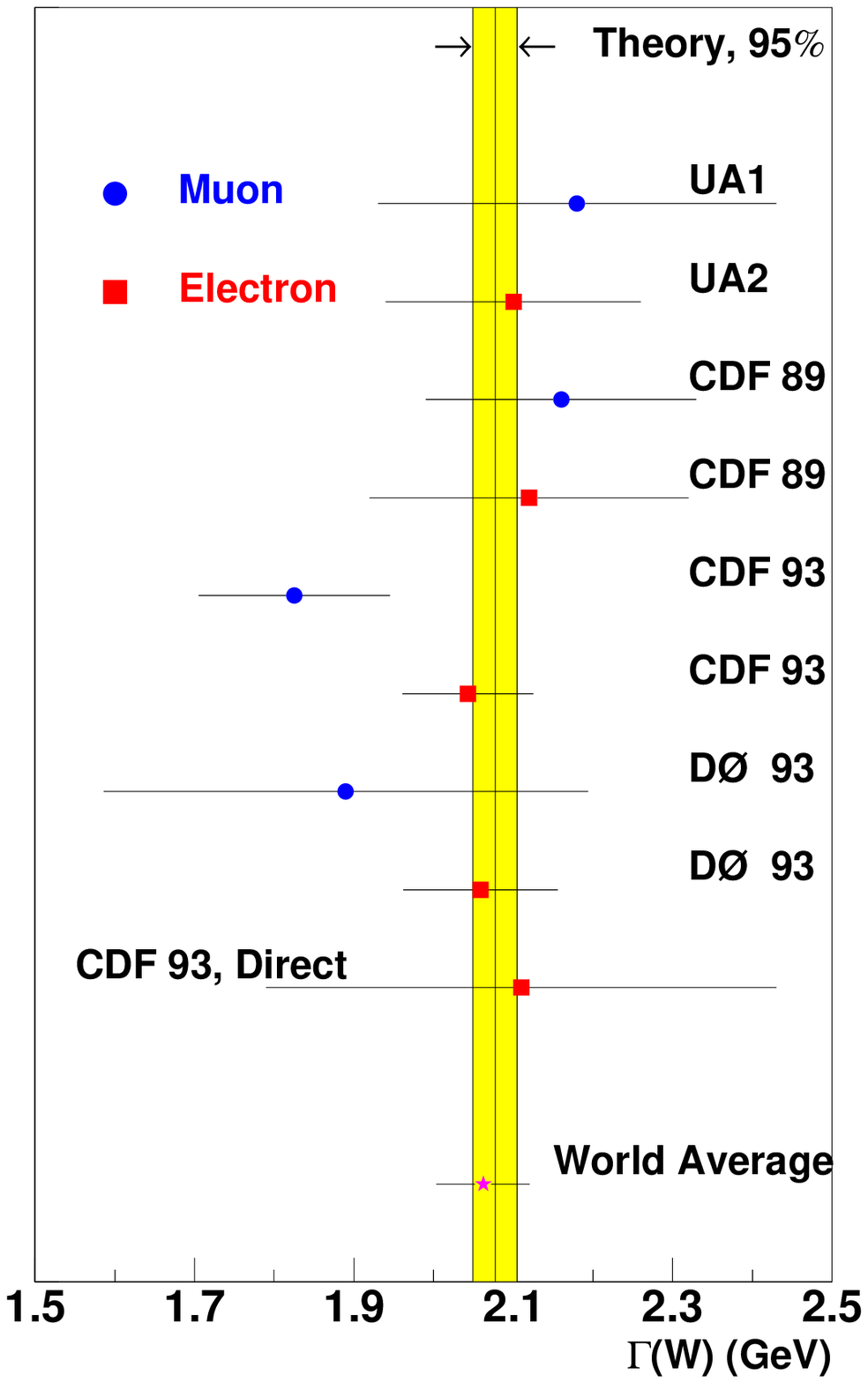,                   height=0.34\textheight,clip=}}
\end{center}
\caption{${\mathrm{Sp\bar{p}S}}$ and Tevatron measurements of the 
         ratio $R$ and of the W width.\label{fig:wwidth}}
\end{figure}

These measurements may alternatively be used to determine 
$\BR{\W\rightarrow \ell \nu}\equiv{\Gamma_{\ell\nu}}/{\GW}$. 
Experimentally, it is appropriate to measure the ratio $R$, defined as
%
%
$R  =  \left( {{\sigma_\W} \cdot \BR{\W\rightarrow \ell \nu}}  \right)/$
$      \left( {{\sigma_\Z} \cdot \BR{\Z\rightarrow \ell \ell}} \right)$
%
%
since a number systematics uncertainties cancel, including 
that from the luminosity determination and   
some of those from the acceptance and efficiency estimations.
Fig.~\ref{fig:wwidth}(left) summarises the measurements of $R$ from
the Tevatron and the ${\mathrm{Sp\bar{p}S}}$ experiments\cite{DEMARTEAU}.
The W leptonic branching ratio is given by 
%
%
$\BR{\W\rightarrow \ell \nu}={R}\times$
$\left({\sigma_\Z} / {\sigma_\W} \right)\times$
${\BR{\Z\rightarrow \ell \ell}}$
%
%
The ratio of cross-sections is taken from the theory\cite{HAMBERG91,NEERVEN92} to be
$\sigma_\W / \sigma_\Z = 3.33\pm0.03$ at ${\sqrt{s}}=1.8$\,TeV and
$\sigma_\W / \sigma_\Z = 3.26\pm0.09$ at ${\sqrt{s}}=0.63$\,TeV,
again with the advantage that some of the systematic uncertainties cancel.
The LEP measurement of $\BR{\Z\rightarrow \ell^+ \ell^-} = 
(3.366 \pm 0.006)$\%\cite{PDG96} is used.
The results from CDF, 
$\BR{\W\rightarrow \ell \nu} = (10.94 \pm 0.45)$\%\cite{DEMARTEAU},
and from \DO, 
$\BR{\W\rightarrow \ell \nu} = (10.43 \pm 0.44)$\%\cite{DARIEN},
are in good agreement with the preliminary measurements from LEP,
the average of which is 
$\BR{\W\rightarrow \ell \nu} = (11.0 \pm 0.7)$\%\cite{GURTU}.

\section{W Width\label{sec:wwidth}}

Precise measurements of the W width, like those of the Z width, may 
yield evidence for non-standard decays involving, for example,
supersymmetric particles\cite{BARGER83,DREES88}
or heavy quarks\cite{ALVAREZ88}.
The ratio $R$ measured by the Tevatron and ${\mathrm{Sp\bar{p}S}}$ 
experiments may be used to determine the W width,  
%
$\GW = \left({1}                / {R}                            \right)\times$
$      \left({\sigma_\W}        / {\sigma_\Z}                    \right)\times$
$      \left({\Gamma_{\ell\nu}} / {\BR{\Z\rightarrow \ell \ell}} \right)$.
%
Taking $\Gamma_{\ell\nu} = (225.2 \pm 1.5)$\,MeV from theory\cite{ROSNER94}, 
$\sigma_\W / \sigma_\Z$ from theory\cite{HAMBERG91,NEERVEN92} as above,
and the LEP\cite{PDG96} measurement of $\BR{\Z\rightarrow \ell^+ \ell^-}$
yields the results for $\GW$ shown in Fig.~\ref{fig:wwidth}(right).
This figure also shows the less-precise but direct measurement by 
CDF of $\GW$ from an analysis of the W transverse mass distribution,
described in section~\ref{sub:wmasstev}.

The world average W width of $\GW = 2.062 \pm 0.059$\,GeV\cite{DEMARTEAU},
which allows for correlations from the uncertainty in the 
PDF's, agrees well with the Standard Model prediction 
of $\GW = 2.077 \pm 0.014$\,GeV\cite{DEMARTEAU}.
The allowed partial width for non-Standard Model 
decays is less than 109\,MeV at the 95\% confidence level.
 
\section{Anomalous Couplings of the W\label{sec:wcouplings}}

Triple Gauge Couplings (TGC's) are central to the Standard Model, for 
example to preserve unitarity for such sets of diagrams such as those 
shown in Fig~\ref{fig:wwfeynman}(upper).
These delicate cancellations are sensitive to physics beyond 
the Standard Model.
The generalised Lorentz-invariant ${\mathrm{WWZ}}$/${\mathrm{WW\gamma}}$ 
vertices have 14 couplings.
The experiments adopt a pragmatic approach and consider only  
$g^{\mathrm{V}}_1$,    
$\kappa_{\mathrm{V}}$,  and
$\lambda_{\mathrm{V}}$ 
$({\mathrm{V}} = \gamma, {\mathrm{Z}})$,
which are theoretically favoured\cite{LEPIIYELLOW}.  
At tree-level, these may be interpreted in terms of the CP conserving quantities: 
electric charge,         $Q_{\mathrm{W}} = e g^{\mathrm{\gamma}}_1$;  
magnetic dipole moment:  $\mu_{\mathrm{W}} = \left({e}/{2\MW}  
                                  \right) (g^{\mathrm{\gamma}}_1  + \kappa_\gamma + \lambda_\gamma)$;  
and electric quadrupole moment:  $q_{\mathrm{W}} = -\left({e}/{\MW^2}\right) 
                                 (\kappa_\gamma - \lambda_\gamma)$.  
There are similar relations, involving also the Weak mixing angle $\theta_{\mathrm{W}}$, 
for $g^{\mathrm{Z}}_1$, $\kappa_{\mathrm{Z}}$,  and $\lambda_{\mathrm{Z}}$. 
The Standard Model predicts: 
$g^\gamma_1         = g^{\mathrm{Z}}_1     = 1$ 
(i.e. $\Delta g^{\mathrm{V}}_1   \equiv  g^{\mathrm{V}}_1 - 1 = 0$),
$\kappa_\gamma   = \kappa_{\mathrm{Z}}  = 1$ 
(i.e.  $\Delta\kappa_{\mathrm{V}} \equiv  \kappa_{\mathrm{V}} - 1 = 0$),  and
$\lambda_\gamma  = \lambda_{\mathrm{Z}} = 0$.
%
\subsection{Anomalous Couplings of the W at LEP}

Anomalous couplings at LEP2 are probed by analysing the differential cross-sections
for W-pair production in terms of the polar angle of the ${\mathrm{W^-}}$,
and the polar and azimuthal angles of the ${\mathrm{W^\pm}}$ decay products
in the ${\mathrm{W^\pm}}$ rest frames.
Despite the low statistics of the samples, the first TGC results are impressive.
For example, an L3 analysis has excluded $g_{\mathrm{ZWW}} = 0$ at $>$ 95\% C.L.

Using a complementary method, L3 has measured the cross-section at ${\sqrt{s}}=172$\,GeV 
for the single W production\cite{SINGLEW} process, ${\mathrm{e^+e^- \rightarrow e\nu_e W}}$,
which is sensitive to $\Delta\kappa_\gamma$ and $\lambda_\gamma$.
The measured cross-section, $\sigma = 0.61 ^{+0.43}_{-0.33} \pm 0.05$\,pb 
is consistent with the Standard Model and therefore the following 95\% C.L.
limits are set:
$-3.6 < \Delta\kappa_\gamma < 1.5$ and 
$-3.6 < \lambda_\gamma      < 3.6$.

The first LEP TGC results are now being finalised for the (later) summer conferences.
In particular they will use a common base set of anomalous couplings 
parameters and include LEP averages allowing for correlations.

\subsection{Anomalous Couplings of the W at the Tevatron}

The Tevatron analyses necessarily allow for a form-factor
dependence of the anomalous couplings:
$\Delta\kappa(\hat{s}) = {\Delta\kappa} / {(1 + \hat{s} / \Lambda^2)^2}$;
and
$\lambda(\hat{s})      = {\lambda}      / {(1 + \hat{s} / \Lambda^2)^2}$;  
where $\hat{s}$ is the effective centre of mass energy of the process and 
$\Lambda$ is the energy scale which is probed for new physics. 

The W pair production cross-section is used to constrain the 
anomalous couplings.
For example, CDF has studied the process 
${\mathrm{p\bar{p}}}\rightarrow {\mathrm{W^+W^-}}\rightarrow \ell\nu\ell\nu$
using $\int{\cal{L}} = 108 {\mathrm{pb}}^{-1}$.  
Di-lepton events (ee/e$\mu$/$\mu\mu$) with $\ETMISS$ are selected with 
high efficiency and low backgrounds.      
The ${\mathrm{t\bar{t}}}$ background is reduced 
by limiting the hadronic (b jet) activity in the event.                   
CDF observes 5 events, with an expected background of $1.2\pm0.3$ events,
which corresponds to a cross section of
$\sigma({\mathrm{p\bar{p}}}\rightarrow {\mathrm{W^+W^-}}) = (10.2^{+6.3}_{-5.1} \pm 1.6)$\,pb.
Since this is in agreement with the Standard Model expectation of 
$\sigma({\mathrm{p\bar{p}}}\rightarrow {\mathrm{W^+W^-}}) = (9.5 \pm 1.0)$\,pb,
they obtain the following constraints, for $\Lambda = 2$\,TeV:
 $-1.0 < \Delta\kappa < 1.3$ (for $\lambda=0$) and  
 $-0.9 < \lambda      < 0.9$ (for $\Delta\kappa=0$)   
assuming $\Delta\kappa_\gamma = \Delta\kappa_{\mathrm{Z}}$ and 
         $\lambda_\gamma      = \lambda_{\mathrm{Z}}$.

The processes
${\mathrm{p\bar{p}}}\rightarrow {\mathrm{WW}}\rightarrow$ jet jet  $\ell\nu$
and 
${\mathrm{p\bar{p}}}\rightarrow {\mathrm{WZ}}\rightarrow$ jet jet $\ell^+\ell^-$ 
have also been analysed, with the advantage that they have 
higher statistics than the ${\mathrm{WW}}\rightarrow\ell\nu\ell\nu$ analyses.
The leptonic W decays are tagged by a high $p_T$ charged lepton and $\ETMISS$,
the hadronic decays by two jets with an invariant mass consistent with $\MW$,
and the Z decays by two charged leptons with an invariant mass consistent with $\MZ$.                       
The large background from W/Z + jets is reduced by requiring the 
W to have high $p_T$.                                  
The sensitivity to SM WW/WZ production is thereby sacrificed 
but not the sensitivity to anomalous processes which tend to populate the 
high $p_T^{\mathrm{W}}$ region.    
For $\Lambda = 2$\,TeV, CDF obtains: 
 $-0.5 < \Delta\kappa < 0.6$ (for $\lambda=0$) and  
 $-0.4 < \lambda      < 0.3$ (for $\Delta\kappa=0$)   
assuming $\Delta\kappa_\gamma = \Delta\kappa_{\mathrm{Z}}$ and 
         $\lambda_\gamma      = \lambda_{\mathrm{Z}}$.
In this analysis, \DO\ enhances the sensitivity by fitting the lepton spectrum
but, since only the Run 1a data is used, less stringent constraints are obtained. 

The radiation of photons from W's is also sensitive to anomalous couplings.
For example, \DO\ uses their standard W sample and requires in addition a 
high transverse energy photon ($E_T^\gamma > 10$\,GeV).
The large background from initial/final state radiation from fermions                      
is suppressed by requiring the photon to be well isolated.
From a fit to the $E_T^\gamma$ spectrum \DO\ obtains, for $\Lambda = 1.5$\,TeV: 
 $-1.0 < \Delta\kappa < 1.0$ (for $\lambda=0$) and  
 $-0.3 < \lambda      < 0.3$ (for $\Delta\kappa=0$)   
assuming $\Delta\kappa_\gamma = \Delta\kappa_{\mathrm{Z}}$ and 
         $\lambda_\gamma      = \lambda_{\mathrm{Z}}$.
%

\section{W Mass\label{sec:wmass}}

At LEP there are two complementary methods for determining the W mass:
from measurements of the W-pair production cross-section close to threshold
and from the direct reconstruction of the decay products of the W's.
The Tevatron experiments use their high statistics W samples to 
perform W mass measurements by direct reconstruction.

\subsection{W Mass from LEP\label{sub:wmasslep}}

\begin{figure}[htbb]
\begin{center}
\fbox{\epsfig{file=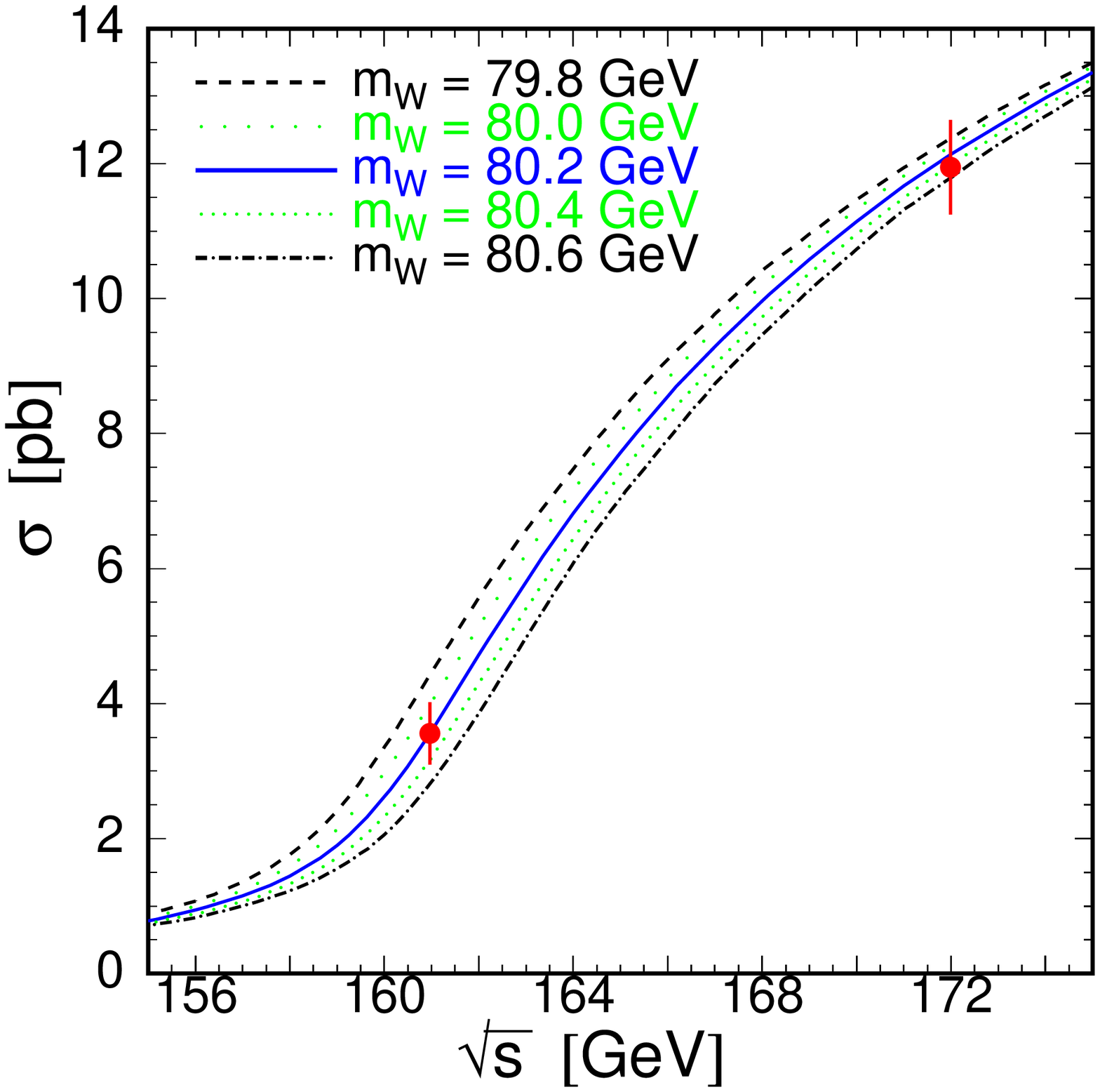,height=0.32\textheight,clip=}} 
~~
\fbox{\epsfig{file=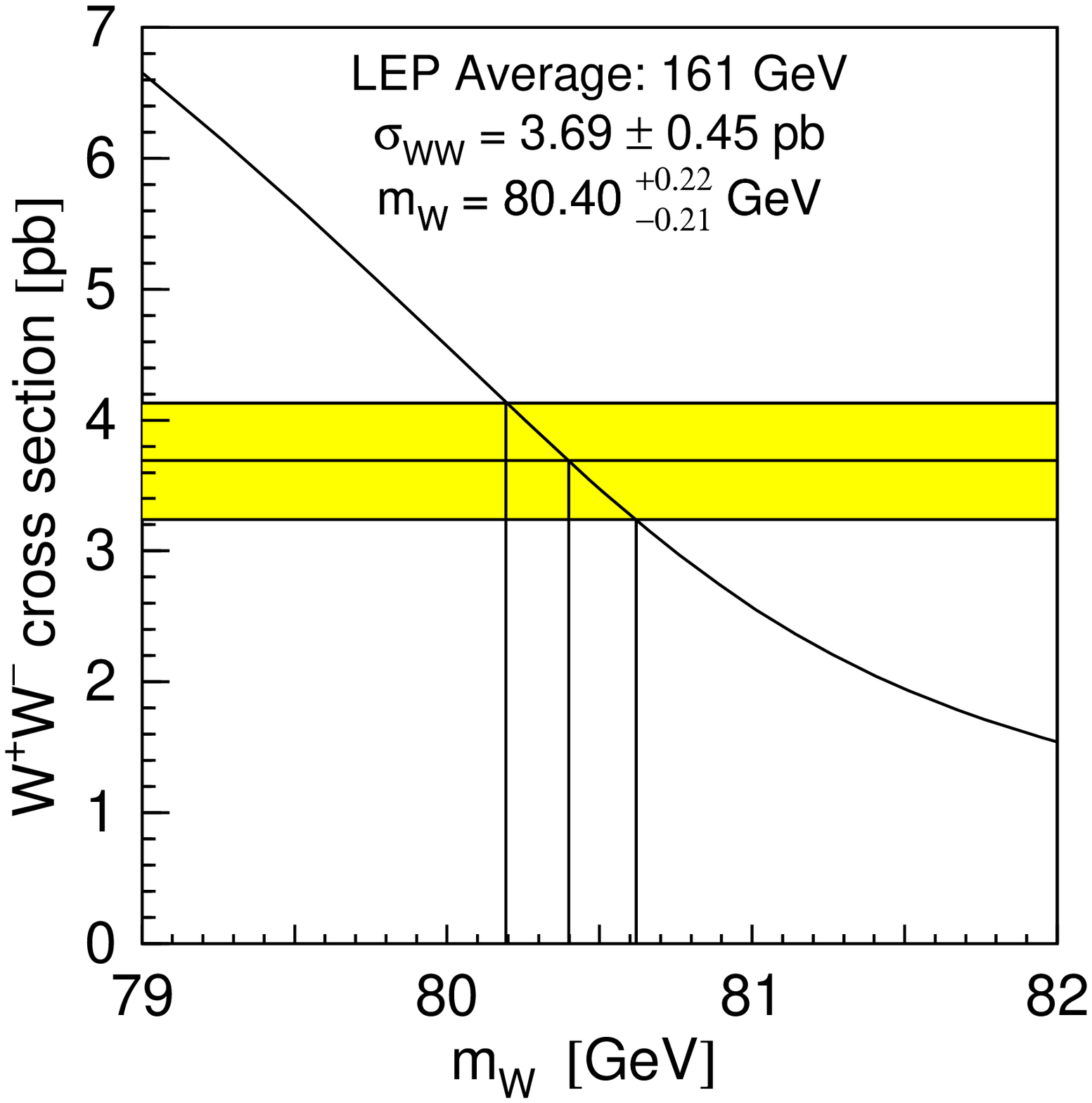,    height=0.32\textheight,clip=}}
\end{center}
\caption{Left: 
         Variation of the W-pair production cross-section in ${\mathrm{e^+e^-}}$
         collisions in the threshold region for various values 
         of the W mass (as predicted by the GENTLE program).
         The data points, which corresponds to the LEP average cross-sections,
         illustrate the sensitivity of a threshold cross-section
         measurement to the W mass. 
         Right:
         The determination of the W mass from the variation of the predicted
         cross-section (at ${\sqrt{s}}=161.33\pm0.05$\,GeV)
         as a function of the W mass.         
         \label{fig:sigthresh}}
\end{figure}

Fig.~\ref{fig:sigthresh}(left) shows the variation of the W-pair production 
cross-section (CC03 processes) in ${\mathrm{e^+e^-}}$ collisions in the 
threshold region for various values of the W mass as predicted in the context of the 
Standard Model by the GENTLE\cite{GENTLE} program.
For the threshold method the optimum sensitivity to \MW\ is at 
$\sqrt{\mathrm{s}} \approx 2\MW + \mathrm{0.5~\GeV}$,
hence the choice of ${\sqrt{s}}=161.33\pm0.05$\,GeV for the initial
phase of LEP running in 1996.       
Fig.~\ref{fig:sigthresh}(right) shows the predicted cross-section as 
a function of \MW, for ${\sqrt{s}}=161.33$\,GeV.
Using the measurement,  
$\sigma_{\mathrm{WW}}(161{\mathrm{GeV}}) = (3.67 \pm 0.42)$\,pb
the LEP average for \MW\ is determined to be\cite{LEPPRIVW}  
$\MW = 80.40 \pm 0.22$\,GeV\OK.
Allowance is made for common systematic errors (0.07\,GeV) which include only 
a small contribution (0.03\,GeV) from the uncertainty in the LEP centre-of-mass energy.
\begin{figure}[htbb]
\begin{center}
\epsfig{file=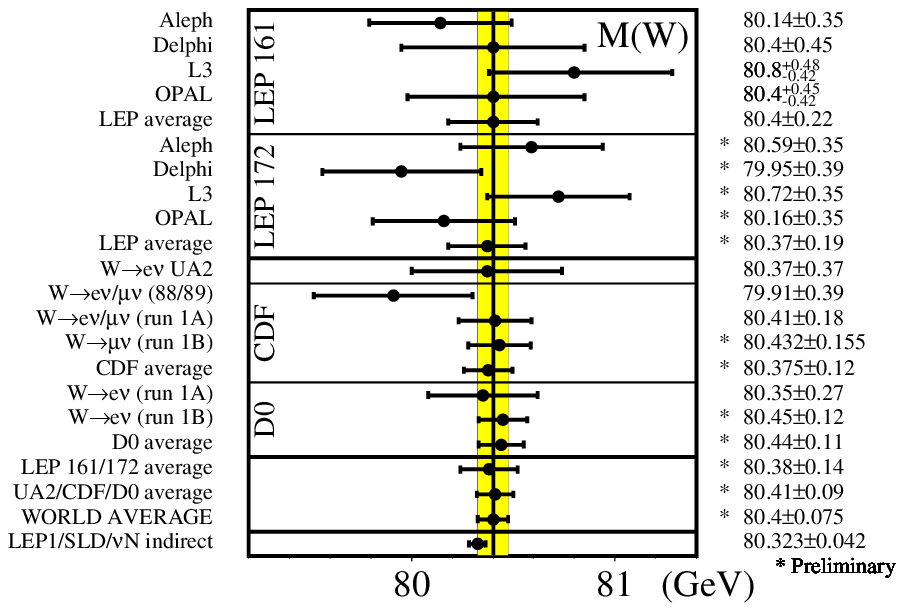,width=0.95\textwidth,clip=}
\end{center}
\caption{Summary of W mass measurements.\label{fig:wmass}}
\end{figure}
The individual LEP measurements are shown in Fig.~\ref{fig:wmass}.

\begin{figure}[!htbb]
\begin{center}
  \epsfig{file=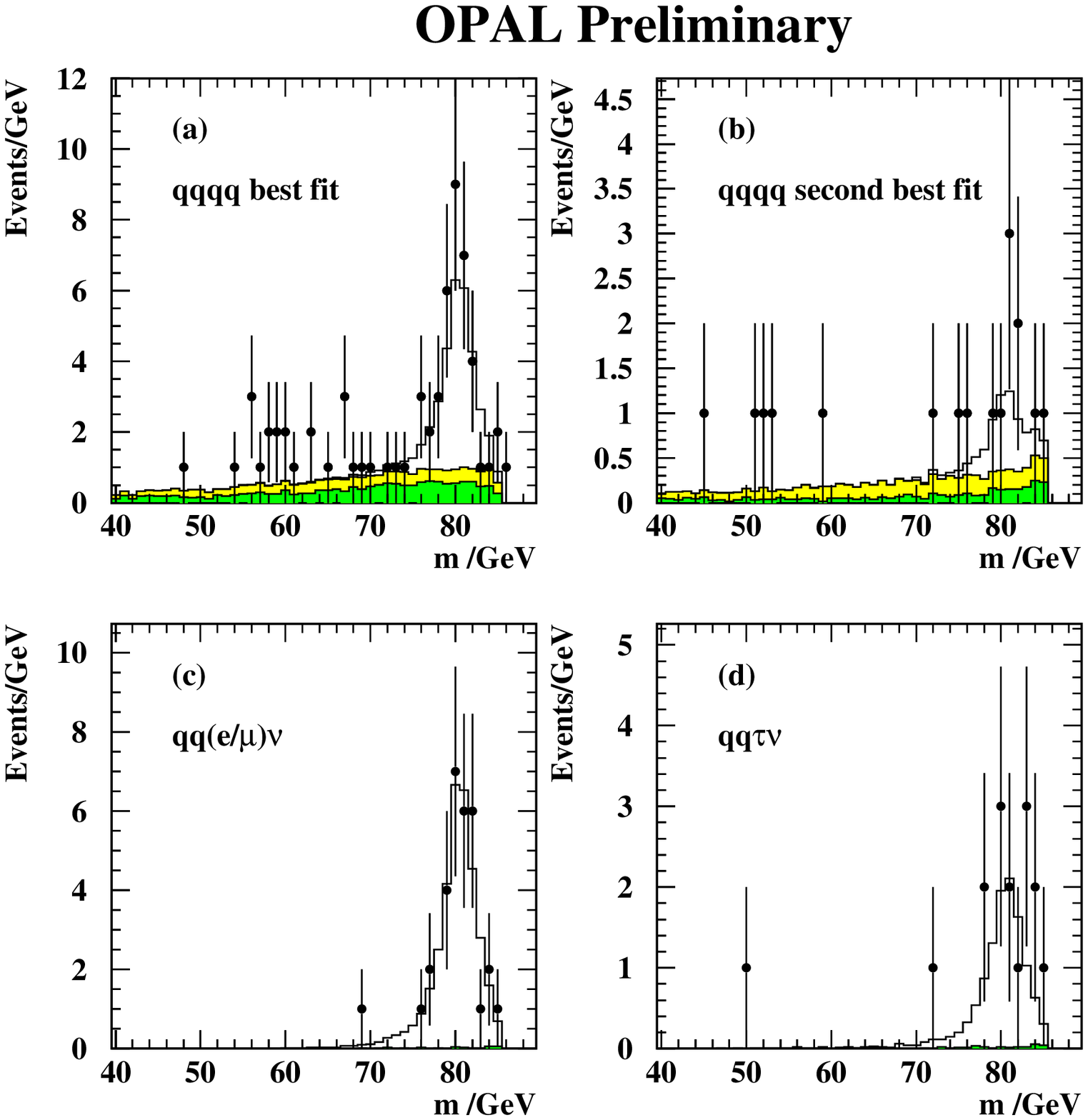,    height=0.6\textheight,clip=}
\end{center}
\caption{Reconstructed invariant mass spectra from the OPAL W-pair analysis 
         at ${\sqrt{s}}=172$\,GeV for 
         (a) best-fit jet-pairing in ${\mathrm{W^+W^-\rightarrow q\bar{q}q\bar{q}}}$ events;
         (b) second-best-fit jet-pairing in ${\mathrm{W^+W^-\rightarrow q\bar{q}q\bar{q}}}$ events;
         (c) ${\mathrm{W^+W^-\rightarrow q\bar{q}\ell\nu~~(\ell=e,\mu)}}$, and 
         (d) ${\mathrm{W^+W^-\rightarrow q\bar{q}\tau\nu}}$.
         The data are shown as points with error bars. 
         The open histogram denotes the total Monte Carlo expectation normalised to 
         the integrated luminosity, 
         the light histogram denotes combinatorial background,
         and the dark histogram denotes other backgrounds.\label{fig:opalwmass}}
\end{figure}

At ${\sqrt{s}}=172$\,GeV the W mass is determined by reconstruction of the W 
decay products using the channels
${\mathrm{W^+W^-\rightarrow q\bar{q}q\bar{q}}}$ and 
${\mathrm{W^+W^-\rightarrow q\bar{q}\ell\nu}}$ 
${\mathrm{(\ell=e,\mu,\tau)}}$\cite{THOMSON}.
For the ${\mathrm{W^+W^-\rightarrow q\bar{q}q\bar{q}}}$ channel,
selected events are forced to contain four jets.
The two correct pairs of jets yield two measurements of the W mass for each event.
There are a number of ways of kinematically reconstructing the W mass 
for a given event: 
using the ``raw'' reconstructed jet pairs; 
scaling the jet energies such that each pair of jets has the beam energy;
performing a 4-C fit by enforcing energy and momentum conservation on the whole event;
performing a 5-C fit where the two measured W masses are forced to be equal.
Typically, the scaling of the jet energies gives a large improvement while the 
additional gains of the 4-C and 5-C fits are more modest. 
There are three possible pairings of jets, only one of which 
corresponds to the two W decays. 
Typically the pair with the highest kinematic fit probability is retained
although some analyses 
also use the second pairing if it has a sufficiently high probability.
Fig.~\ref{fig:opalwmass} shows the results of the OPAL W mass analysis
of the ${\mathrm{W^+W^-\rightarrow q\bar{q}q\bar{q}}}$ channel for
(a) the best fit pairing (60 events expected and 67 observed), 
and 
(b) the second best fit pairing (17 events expected and 17 observed).

The ${\mathrm{W^+W^-\rightarrow q\bar{q}\ell\nu}}$ analyses force the selected 
events to contain two jets.
The charged lepton is reconstructed with high precision and efficiency using standard 
techniques developed at LEP 1, while the neutrino kinematics are inferred 
from the missing energy in the event.
This channel has no combinatorics and less background than the 
four-jet channel.
However, since reconstruction of the neutrino ``uses up'' three degrees of freedom, 
only 1-C or 2-C fits may be used for the extraction of the W mass. 
Fig.~\ref{fig:opalwmass} shows the results of the OPAL W mass analysis
for (c) ${\mathrm{W^+W^-\rightarrow q\bar{q}(e/\mu)\nu}}$ (31 events seen for 33 expected) and 
    (d) ${\mathrm{W^+W^-\rightarrow q\bar{q}\tau\nu}}$ (16 events seen for 12 expected).

The W mass is obtained from the data assuming that the 
invariant mass distribution is described 
by a relativistic Breit-Wigner convoluted with an experimental 
resolution function and a phase-space function to account for the fit constraints.
Biases from the following sources are also taken into account:
initial state radiation, 
the event selection efficiencies,
the mass reconstruction, 
and the fitting method.
These are generally determined from Monte Carlo studies of the differences in the 
true and fitted W masses and applied to the 
value of \MW\ obtained from the data.
In the L3 analysis, such effects are implicitly included through the use 
of many Monte Carlo samples (with differing input W masses) which are used 
to construct the likelihood distribution for the data as a function of \MW.

Fig.~\ref{fig:wmass}\OK\ summarises 
the W mass measurements\cite{THOMSON,LEPPRIVW} from ALEPH, DELPHI, L3, and 
OPAL using the data at ${\sqrt{s}}=172$\,GeV.
%
The systematic errors on \MW, using OPAL as an example\cite{THOMSON}, include:
detector effects            (64\,MeV)\OK;
hadronisation               (44\,MeV)\OK;
fit procedure               (43\,MeV)\OK;
initial state radiation     (27\,MeV)\OK;
colour recombination and 
Bose-Einstein correlations  (50\,MeV)\OK;
LEP beam energy uncertainty (30\,MeV)\OK.
The latter two are completely correlated between experiments and are taken 
into account when deriving the average LEP 172 value~\cite{LEPPRIVW} of
$\MW = 80.37 \pm 0.19$\,GeV\OK.
Many of these systematic errors will be reduced as more data are collected 
and analysed although future reductions in the uncertainty from the colour recombination and 
Bose-Einstein correlations will require theoretical as well as 
experimental input.

The average of all the LEP threshold and the direct reconstruction measurements is
$\MW = 80.38 \pm 0.14$\,GeV\OK, where allowance has been made 
for common systematic errors\cite{LEPPRIVW}.

\subsection{W Mass from the Tevatron\label{sub:wmasstev}}

The CDF W mass measurements are based on 
24\,pb$^{-1}$\OK of ${\mathrm{W\rightarrow({e}/\mu)\nu}}$ data from 1988/89\cite{WMASSCDFA,WMASSCDFB}
and Run 1a\cite{WMASSCDFC,WMASSCDFD} 
and, more significantly, on $\sim 90$\,pb$^{-1}$\OK of ${\mathrm{W\rightarrow\mu\nu}}$ data from
Run 1b\cite{GORDON,WMASSCDFE}.
\DO\ W mass measurements are based on ${\mathrm{W\rightarrow{e}\nu}}$ data from
Run 1a\cite{WMASSDZEROA}        ($\sim$15\,pb$^{-1}$) and 
Run 1b\cite{DARIEN,WMASSDZEROB} ($\sim$80\,pb$^{-1}$).

W mass measurements from the Tevatron (see review of M. Demarteau\cite{DEMARTEAU} for more details)
are based on fits to distributions of transverse mass, 
$m_T = {\sqrt{2 p_T^\ell  p_T^\nu (1 - \cos \phi^{\ell\nu})}}$,
where:
$p_T^\ell$ and $p_T^\nu$ denote the transverse momenta of the 
charged lepton (e or $\mu$) and the neutrino respectively,
and $\phi^{\ell\nu}$ denotes the angle between the charged lepton
and the neutrino in the transverse plane.
The value of $p_T^\nu$ is inferred from the measured missing energy, 
allowing for the luminosity-dependent transverse energy flow 
of the underlying minimum-bias interactions.
The transverse mass is preferred to the transverse momentum of 
the charged lepton because to first order it is independent of 
the modelling of the transverse momentum of the W.
On the other hand, it relies on the precise and accurate determination 
of $p_T^\nu$ which is a challenging experimental task.

For the W mass measurement it is crucial to understand the 
energy scale and resolution of the charge lepton measurements.
CDF determines the muon momentum scale using measurements of the
${\mathrm{J}}/\psi$ invariant mass.
The extrapolation from the ${\mathrm{J}}/\psi$ to the higher average momenta 
of ${\mathrm{W\rightarrow\mu\nu}}$ decays, allowing for possible non-linearities,
is cross-checked using measurements of the $\Upsilon$ and Z invariant masses. 
The uncertainty on the muon momentum scale results in a 40\,MeV uncertainty
on the W mass measurement. 
\DO\ has performed a similar study to determine the electron energy scale,
using the measured invariant mass spectra for
$\pi^0\rightarrow\gamma\gamma$, 
${\mathrm{J}}/\psi\rightarrow{\mathrm{e^+e^-}}$, 
and ${\mathrm{Z}}\rightarrow{\mathrm{e^+e^-}}$ decays.
Possible non-linear effects are constrained using test-beam data.
The charged lepton energy resolutions are verified using the 
${\mathrm{Z}}\rightarrow{\mathrm{\ell^+\ell^-}}$ samples.

\begin{figure}[!tb]
\parbox[b]{0.5\textwidth}
{%
\epsfig{file=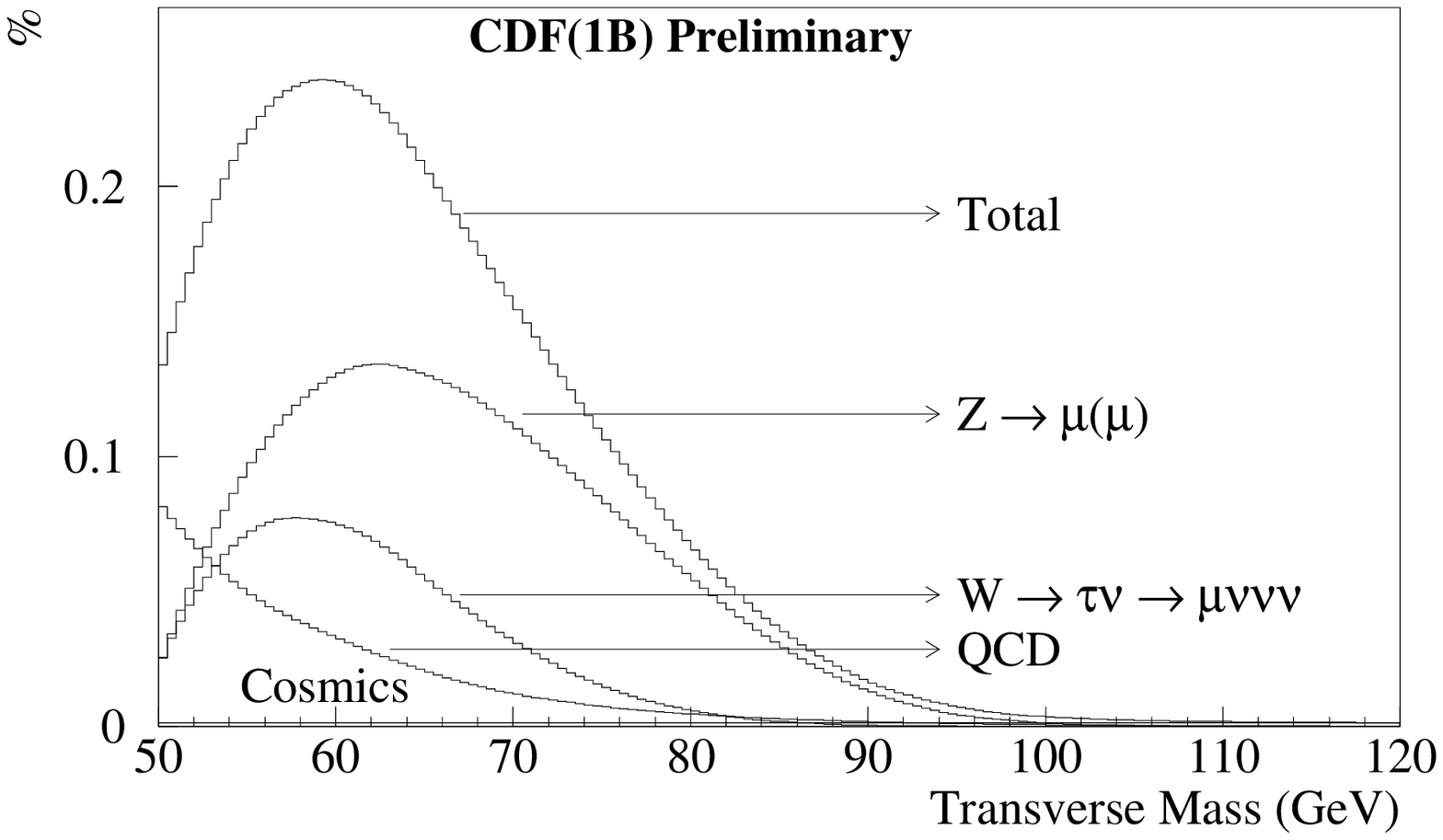,width=0.475\textwidth,clip=}\\%
\epsfig{file=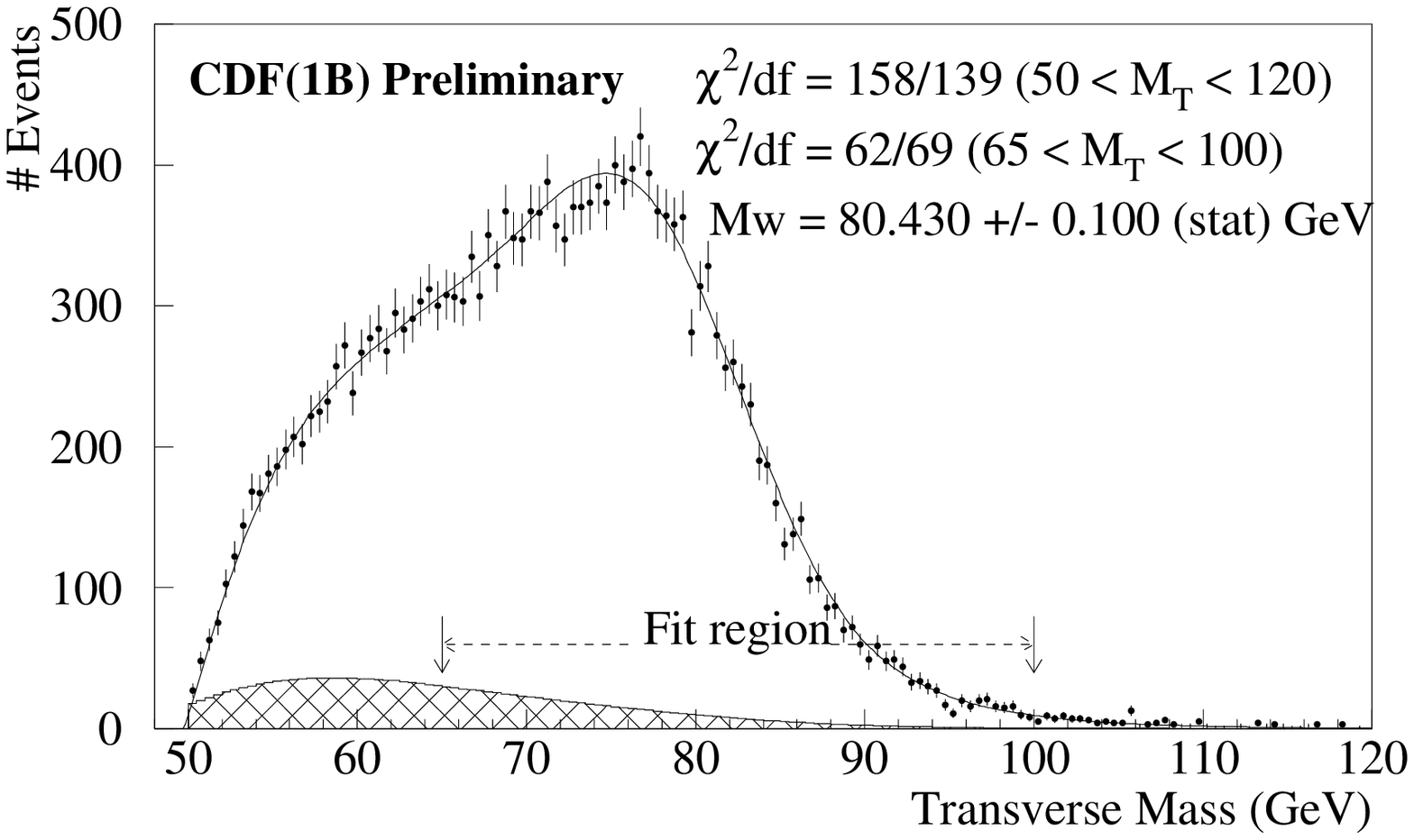,width=0.475\textwidth,clip=}%
}
\parbox[b]{0.5\textwidth}
{%
\epsfig{file=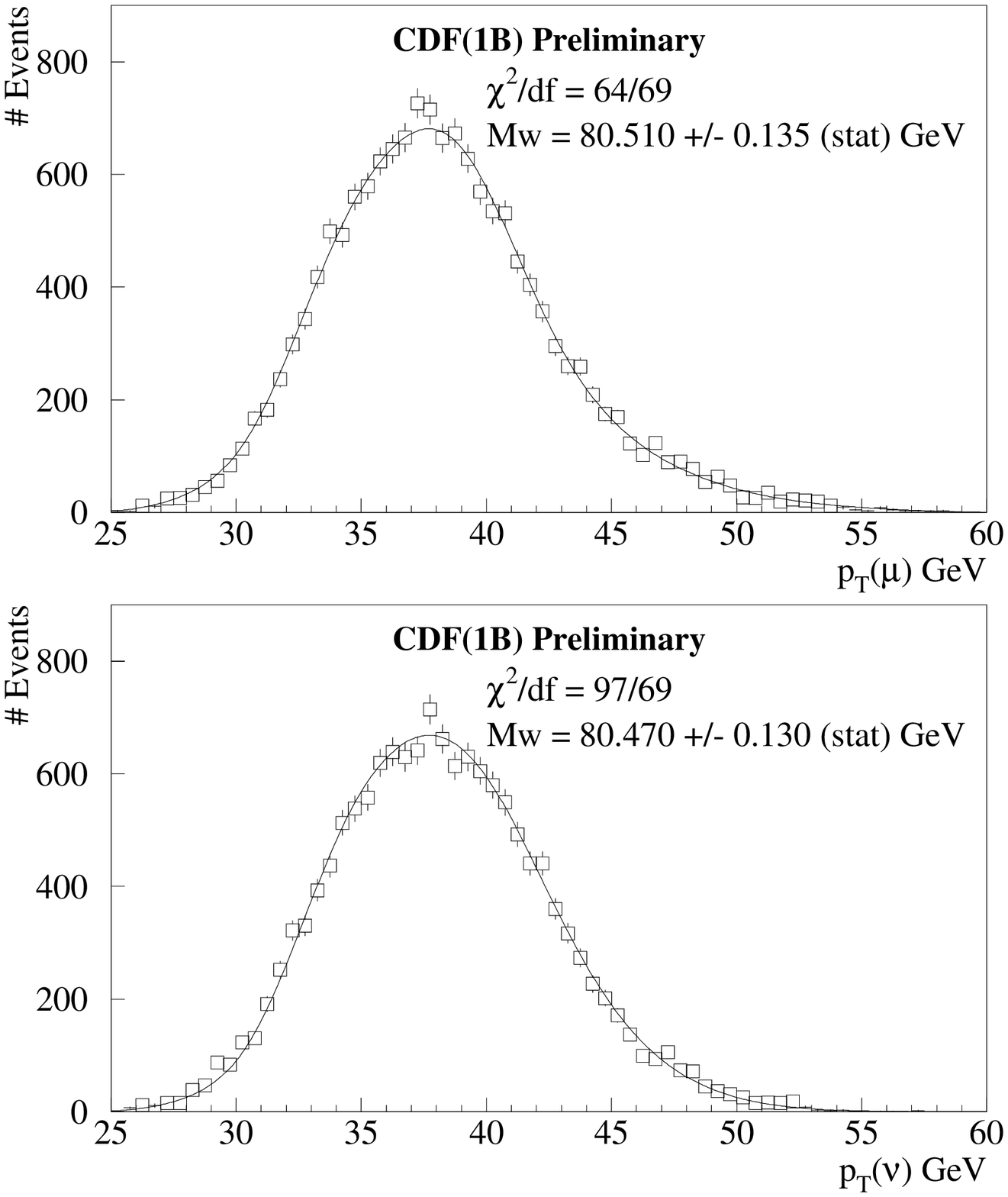,  width=0.48\textwidth,clip=}%
}
\caption{Distribution of transverse mass, $m_T$, from the CDF run 1b
         ${\mathrm{W\rightarrow\mu\nu}}$ analysis, for background (upper left)
         and the data (lower left), showing the result of the fit for $\MW$ (solid
         line) and the estimated background contribution (hatched). 
         Distributions of muon (upper right) and neutrino (lower right) 
         transverse momenta from the CDF run 1b
         ${\mathrm{W\rightarrow\mu\nu}}$ analysis, with the results 
         of the fits for $\MW$.\label{fig:cdfmass}}
\end{figure}

The W mass is determined from the measured $m_T$ distributions by comparison with
Monte Carlo distributions with differing input W masses.
The Monte Carlo programs generate the W's as a relativistic Breit-Wigner resonance 
with a longitudinal momentum distribution according to various PDF models
(MRSA$^\prime$, MRSD$^\prime$, CTEQ2M, and CTEQ3M). 
The modelling of the W transverse momentum, 
the underlying event, and multiple interactions
is determined and/or checked using W, Z, and minimum-bias data samples.
Background contributions, which are at the few percent level, include 
${\mathrm{Z}}\rightarrow\ell^+\ell^+$ decays in which one of the leptons is lost,
${\mathrm{W}}\rightarrow\tau\nu; \tau\rightarrow\ell\nu\nu$ decays,
and mis-identified QCD di-jet events.
Fig.~\ref{fig:cdfmass}(upper left) shows the $m_T$ distributions for the 
background contributions to the CDF Run 1b ${\mathrm{W\rightarrow\mu\nu}}$  
analysis.
Fig.~\ref{fig:cdfmass}(lower left) shows the ${\mathrm{W\rightarrow\mu\nu}}$ 
data of CDF together with the result of the fit.
The uncertainties on the CDF W mass from this analysis\cite{WMASSCDFE},
which are typical of the Tevatron analyses, include
the following sources:  
  statistics (100\,MeV); 
  momentum scale (40\,MeV); 
  momentum resolution (25\,MeV); 
  PDF's (25\,MeV); 
  QED (20\,MeV); 
  QCD (20\,MeV); 
  input $p_T$ of the W's (40\,MeV); 
  recoil model (90\,MeV); 
  trigger bias (15\,MeV); 
  selection bias (10\,MeV); 
  background (25\,MeV); and
  fitting systematics (10\,MeV).
As a cross-check, both CDF and \DO\ also fit the charged lepton and 
neutrino $p_T$ distributions to extract the W mass.
They obtain results which are consistent with the $m_T$ fit results,
as shown in Fig.~\ref{fig:cdfmass}(right) for CDF.

The average CDF W mass, allowing for a common error of 60\,MeV for all their 
measurements, is $\MW = 80.375 \pm 0.120$\,GeV\OK\cite{WMASSCDFE}.
After inclusion of the electron channel, CDF anticipates that the 
final W mass error from the Run 1 data will be  
$\sigma(\MW)\approx$100\,MeV\cite{GORDON}\OK.
The average \DO\ W mass is 
$\MW = 80.44 \pm 0.11$\,GeV\OK\cite{WMASSDZEROB}.
\DO\ anticipates that their final Run 1 W mass analysis, including the 
end cap region $(1.5<|\eta_{\mathrm{e}}|<2.5)$, will 
yield a W mass error of $\sigma(\MW)\approx$100\,MeV\cite{DARIEN}.
Fig.~\ref{fig:wmass}\OK\ summarises the W mass measurements from CDF and \DO.
The UA2 result also shown is determined from their measurement of the ratio 
$\MW/\MZ = 0.8813 \pm 0.0036 \pm 0.0019$\cite{WMASSUATWO}\OK, which is
scaled to the LEP value of $\MZ = 91.187 \pm 0.007$\,GeV\cite{PDG96}.
The hadron collider average is $\MW = 80.41 \pm 0.09$\,GeV\OK, where a 
common error of 65\,MeV has been assumed\cite{WMASSCDFE}.

\subsection{Interpretation of W Mass Measurements}

Fig.~\ref{fig:wmass}\OK\ summarises the W mass measurements from 
ALEPH, DELPHI, L3, OPAL, UA2, CDF, and \DO.
The average of these direct measurements is
$\MW = 80.400 \pm 0.075$\,GeV\OK.
This is in good agreement with the indirect prediction from 
a fit to electroweak measurements\cite{CLARKE} 
($\MZ$, $A_{\mathrm{FB}}$, $A_{\mathrm{LR}}$, etc.) of the  
LEP1, SLC, and $\nu$N scattering\cite{CCFR} experiments, which
yields $\MW = 80.323 \pm 0.042$\,GeV\cite{LEPPRIVW}.
The fit also results in predictions for the top and Higgs masses,
as shown in table~\ref{tab:masssum}.
\begin{table}[tbb]
\caption{Indirect and direct determinations of \MW, \MT, and \MH.\label{tab:masssum}}
\vspace{0.4cm}
{\small\renewcommand{\arraystretch}{1.2}
\begin{center}
\begin{tabular}{|l|c|c|c|} \hline
                                           &        $\MW$ (GeV)         &       $\MT$ (GeV)      &        $\MH$ (GeV)          \\ \hline
LEP1/SLC/$\nu$N (indirect)                 &        $80.323 \pm 0.042$  &       $155^{+10}_{-9}$ &     $36^{+52}_{-18}$        \\ \hline 
LEP2/Tevatron  (direct)                    &        $80.400 \pm 0.075$  &       $175.6 \pm 5.5$  &  $>77$ (unofficial)         \\ \hline
Combined fit                               &        $80.366 \pm 0.031$  &       $172.7 \pm 5.4$  &        $127^{+127}_{-72}$   \\ \hline
%
%
\end{tabular}
\end{center}
}
\end{table}
The indirect determination of $\MT$ is in agreement with the direct measurement 
of CDF/\DO\ of $\MT = 175.6 \pm 5.5$\,GeV\cite{RAJA}.

\begin{figure}[!htb]
\begin{center}
\fbox{\epsfig{file=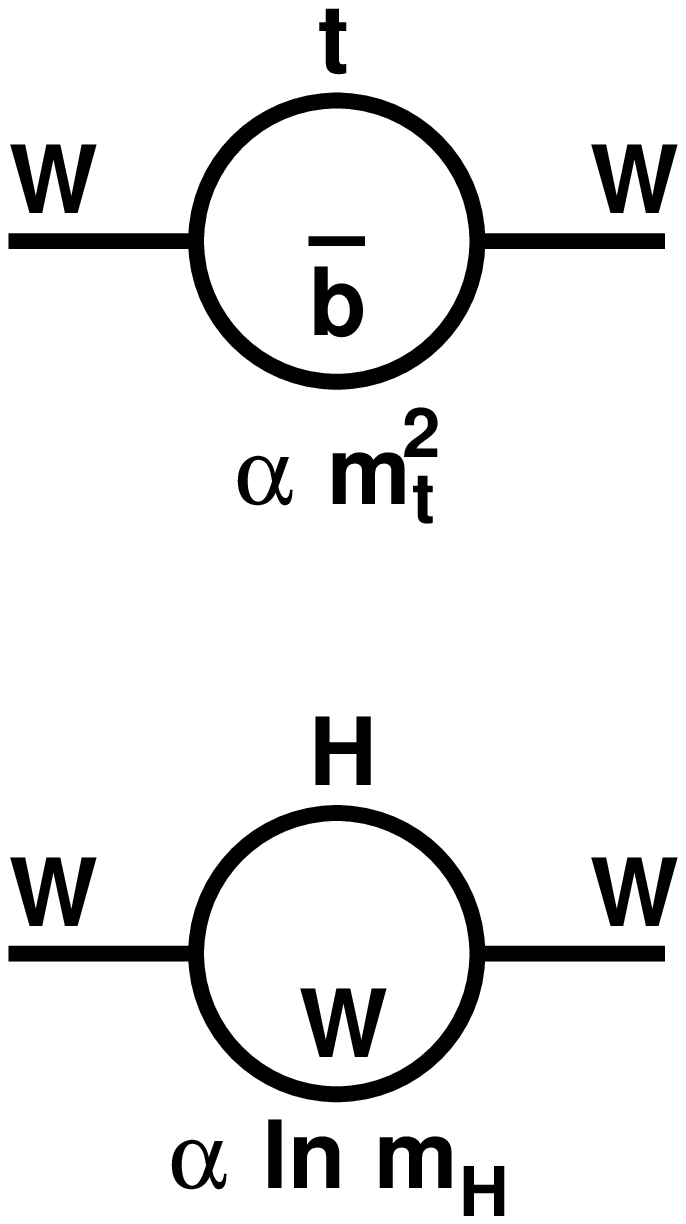,height=0.42\textheight,clip=}} 
~~
\fbox{\epsfig{file=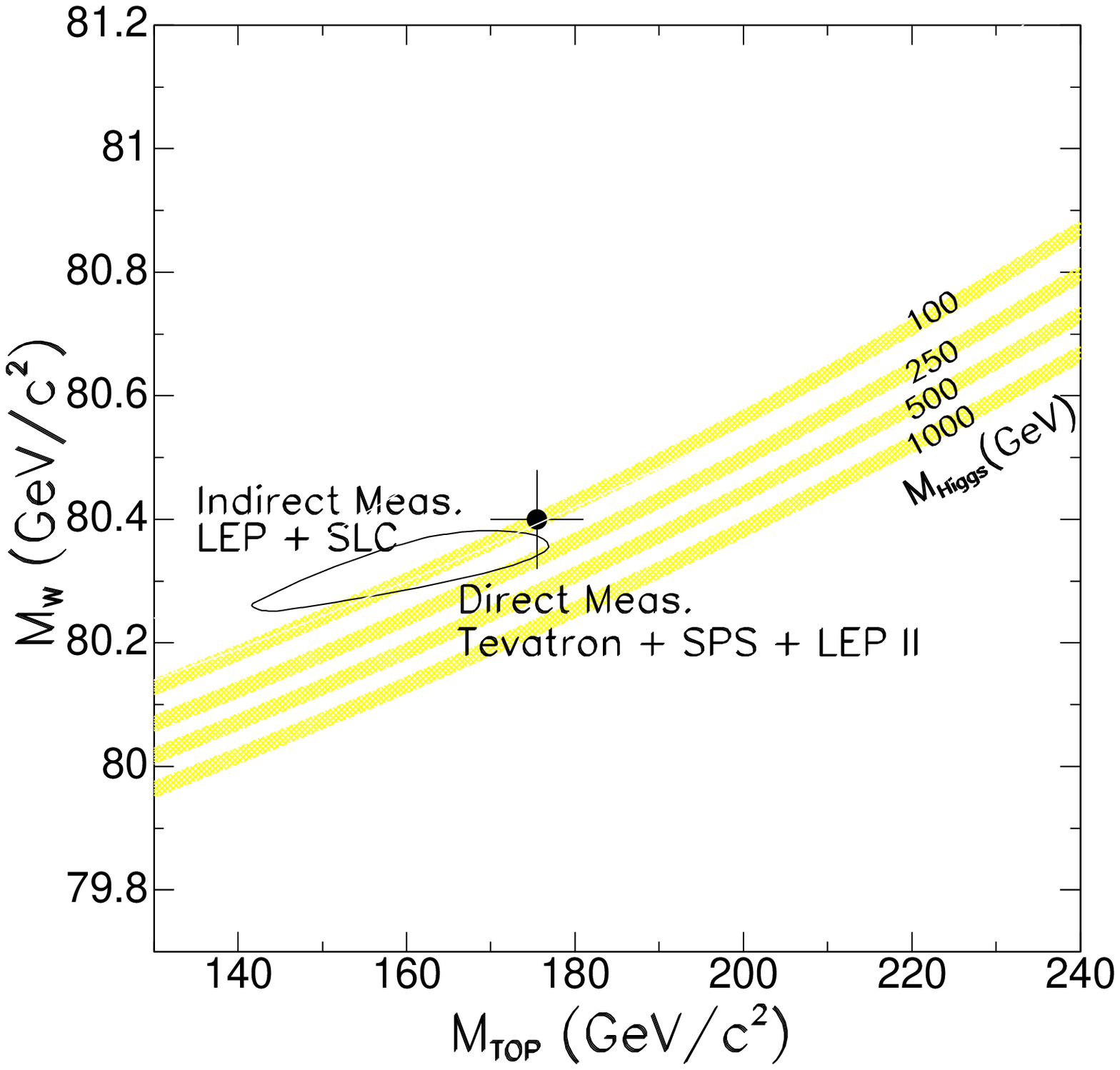,height=0.42\textheight,clip=}}
\end{center}
\caption{Left: Loop corrections giving rise to a dependence of $\MW$ on 
         $\MT$ and $\MH$.  
         Right: The variation of $\MW$ with $\MT$ for various values of $\MH$,
         shown as bands, compared to the direct measurements of 
         $\MW$ from LEP and the Tevatron and of $\MT$ from the Tevatron.
         The LEP/SLC contour for the indirect determination 
         of $\MW$ and $\MT$ is consistent with the direct measurement.\label{fig:higgs}}
\end{figure}

Loop corrections give rise to a quadratic dependence of $\MW$ 
on $\MT$ and a logarithmic dependence of $\MW$ on $\MH$,
as shown schematically in Fig.~\ref{fig:higgs}(left).
Fig.~\ref{fig:higgs}(right) illustrates the variation of $\MW$ with $\MT$ 
for various values of $\MH$, shown as bands.
The data point denotes the direct measurements of $\MW$ from LEP and 
the Tevatron and of $\MT$ from the Tevatron. 
The LEP/SLC/$\nu{\mathrm{N}}$ contour for the indirect determination 
is consistent with the direct measurements. 

A combined fit, in the context of the Standard Model, to the indirect 
electroweak measurements and the direct measurements 
of $\MW$ and $\MT$ yields 
$\MW = 80.366 \pm 0.031$\,GeV, 
$\MT = 172.7  \pm 5.4$\,GeV and 
$\MH = 127^{+127}_{-72}$\,GeV.
The apparently high significance of the low value of $\MH$ should be interpreted with caution;
the likelihood is approximately parabolic in $\log\MH$, therefore the upper bound 
is less constrained than it would appear to be from the quoted $1\sigma$ error.
The upper limit on the Higgs mass at the 95\% confidence level 
is $\MH < 465$\,GeV, indicating that the 
data weakly favour a low mass Higgs.


\section*{Acknowledgements}
I would like to thank all the people who, sometimes unknowingly, greatly 
facilitated the preparation of this review, in particular:
A. Blondel,
S. Christen,
R. Clare,
M. Demarteau,
D. Errede,
P. Fisher,
M. Gr{\"{u}}newald,
A. Gurtu,
J. Mnich,
K. M{\"{o}}nig,
T. Paul,
C. Paus,
D. Stickland,
N. Watson, 
and 
D. Wood.
It is a pleasure to acknowledge the hard work of the organisers
who made the conference such a success.


\end{document}